# How To Build A Superbeam

J. Hylen

*Fermi National Accelerator Laboratory, P.O. Box 500, Batavia, IL 60510, USA*

**Abstract.** A discussion of design issues for future conventional neutrino beam-lines with proton beam power above a megawatt.

## INTRODUCTION

There are several conventional neutrino beam-lines currently in operation that are designed to handle proton beam power of a fraction of a mega-watt. By conventional neutrino beam-line is meant one where accelerated protons strike a target to produce charged pions, which are magnetically focused and allowed to decay to neutrinos. Several laboratories are considering accelerator upgrades over the next decade that could provide proton beam power above a mega-watt for neutrino beam-lines (see Table 1); conventional neutrino beams at such high power have been labeled Superbeams. Based on current experience, the most significant technical issues for this next generation of high-power neutrino beam-lines are radiation protection, target survivability, and reliability/reparability. A few examples are given of extrapolating from NuMI to LBNE (the proposed beam-line from FNAL to DUSEL in South Dakota using protons from the Project X accelerator upgrade).

**TABLE 1.** Possible evolution of proton beam power over the next decade, with beam parameters for Superbeams. For T2K, the neutrino beam-line is ready for 0.75 MW, but accelerator upgrades are needed to reach that power.

|  | CERN | FNAL | JPARC |
|---|---|---|---|
| Operational Neutrino Beam | CNGS (0.5 MW) | NuMI/MINOS (0.4 MW) | T2K (0.1 MW / 0.75 MW) |
| Upgrades by ~ 2013 |  | NuMI/NOVA (0.7MW) | T2K (0.75 MW) |
| Under consideration | SPL to new beam-line | Project X to DUSEL | T2K Roadmap plan |
| Proton beam power | 4 MW | 2.3 MW or 2.0 MW | 1.7 MW |
| Proton Energy | 3.5 GeV | 120 GeV or 60 GeV | 30 GeV |
| Repetition Rate | 50 Hz | 0.75 Hz or 1.3 Hz | 0.5 Hz |
| Protons per spill | $1.4 \times 10^{14}$ | $1.6 \times 10^{14}$ | $6.7 \times 10^{14}$ |

## SUPERBEAM CHALLENGES

We now have substantial experience with conventional neutrino beam-line technology that we can use to extrapolate to higher power. It is important to note that the fixed infrastructure of the T2K beam-line (shielding, cooling and absorber) has already been built to accommodate up to 4 MW beam power, although some replaceable components like the target are designed for 0.75 MW. There do not appear to be any show-stoppers for Superbeams, but there are issues that can lead to qualitative differences in design choices.

The Superbeams will be more expensive and higher profile beam-lines, and one should consider carefully before taking the same level of risk as in previous beam-lines with non-repairable systems. Specifically, what happens if the decay pipe cooling or absorber cooling fails? This could turn the entire facility off forever. So is it worth some extra up-front cost and some loss of neutrino production efficiency to instead build systems that are repairable?

Target design is problematic due to (i) worse stress waves from the fast beam spill (the numbers of protons per spill in Table 1 are 4 to 16 times what the current NuMI target has to take), (ii) higher average thermal load (possibly eliminating options like helium cooling), and (iii) faster radiation damage. These same considerations also apply to beam window design.

The primary beam can do substantially more damage in a single pulse, so minimizing/eliminating stray beam pulses becomes a higher priority.

Residual radiation levels cross the point where hands-on repair becomes impossible, and more emphasis on remote handling is required. Several component repairs of NuMI horn and target systems were accomplished by splitting a simple job among ~10 people, each taking a ~ 10 second radiation dose. Extrapolating this to 100 people at 1 second each at a Superbeam will not work.

The increased heat load implies the target pile shielding probably requires water cooling rather than air cooling.

Air near the target becomes highly corrosive, and materials in contact with the air need to be very carefully selected, or else one needs to fill the entire area with inert gas (as T2K has chosen to do).

A significant challenge is to not spend an order of magnitude more money on a facility that has to take an order of magnitude more beam power.

## SELECTION OF FOCUSING SYSTEM

Focusing system technologies that have been used or considered for conventional neutrino beams include horns, magnetic spokes, solenoid, quadrupole triplet, lithium lens, plasma lens, hadron hose, and dichromatic beam-line utilizing dipole magnets for pion and kaon momentum selection. The neutrino detector can be placed on the magnetic focusing axis for a broad energy spectrum, or off-axis by a few degrees for a narrower energy spectrum at lower energy. Most of these schemes are reviewed in [1].

An off-axis horn focusing system was chosen as optimal for T2K, where one is looking only at the first neutrino oscillation maximum.

For LBNE, the distance to the far detector is selected to be >1000 km to maximize matter effects on neutrino oscillations; the distance from FNAL to DUSEL is 1290 km. LBNE wants to study both the first and second oscillation maxima, which at this distance occur at $E_\nu$ = 2.7 GeV and 0.8 GeV. The need for a high statistics beam covering this broad energy range is matched by solenoid focusing or horn focusing with detector on-axis. Since the 1st round LBNE detectors will not have the ability to separate neutrinos from anti-neutrinos, the sign-selection must be done by the beam, which is natural for the horn system but not the solenoid.

## TARGET SURVIVABILITY

Neutrino beam-line targets are generally about two interaction lengths long so that most of the protons interact, and narrow so that pions can exit out the sides without being re-absorbed. The resulting need for a small beam spot size leads to stress and thermal loading significantly limiting the choice of target material; CNGS, NuMI and T2K all use varieties of graphite.

At LBNE, the pions which contribute to the desired portion of the neutrino spectrum emerge from the target at relatively large angles, (for on-axis beams, the pion angle from the target is ~ 0.1 GeV / $E_\nu$), implying that the target will need to be essentially inside the first horn. The effect of pion re-absorption when increasing the target radius is illustrated in Figure 1.

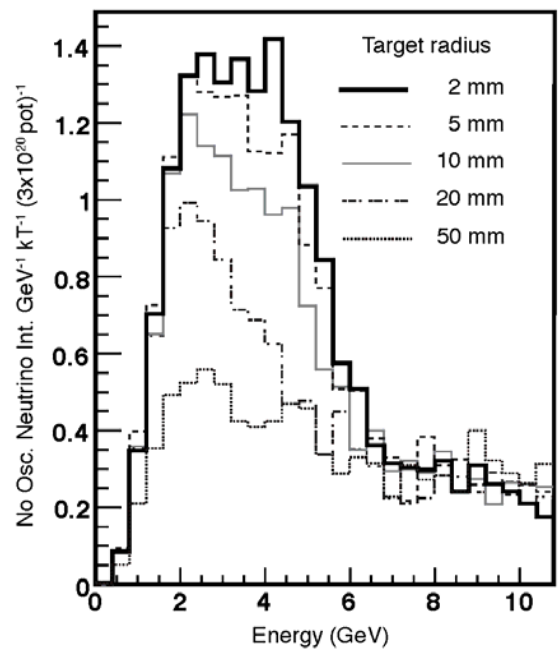

**FIGURE 1.** Expected neutrino interaction rate in LBNE detector as a function of target radius. A graphite target and three horn system similar to that used by T2K is assumed for this calculation.

However, the smaller the target radius and beam spot size the worse the thermal stress from the fast beam spill, the faster the radiation damage, and the harder the problem of protecting the cooling system from errant beam pulses.

A general survey of target issues is beyond the scope of this paper, and the selection of target material, geometry and cooling system optimized for each Superbeam will require significant work. However, extrapolating the NuMI target experience to a fairly similar target for use at LBNE illustrates some of the issues involved.

## Target Stress

The NuMI graphite target consists of 6.4 mm wide, 15 mm high fins, with water cooling tubes soldered to the top and bottom. It has a design safety factor of 1.6 against failure from the stress generated by the short beam spill of $4\times10^{13}$ protons per pulse, and thus would be predicted to fail with the four times as many protons per spill at LBNE. The NuMI beam spot size is 1.1 mm RMS. Increasing the beam spot size to 1.5 mm for LBNE is sufficient to keep the compressive stress at the center of the target under control. However, the stress on the outer corners of the target also grows, and because graphite is much stronger in compression than tension, this becomes the failure point. This can be mitigated by pre-tensioning a metal tube around the graphite; modeling of a rod target within a 15 mm I.D. metal tube shows acceptable stress levels everywhere.

## Target Cooling

For the current generation of targets, CNGS uses radiative and helium convection cooling of the target segments with forced air around the outer canister, NuMI uses water cooling, and T2K uses forced helium flow cooling.

For LBNE, all of the above are problematic. For water flow cooling à la NuMI, the problem is severe water hammer induced by beam heating of the water during the short beam pulse. While there is some hope that a properly engineered cooling pipe will survive the water hammer, several water-cooling variations have been proposed. These include a two-phase water system (for instance a heat pipe where water turns to steam, or helium bubbles entrained in the water), or contact cooling to the horn inner conductor (which is cooled by water-spray). If the target did not need to fit inside the horn, direct water spray on the target tube would be an easy, effective solution.

## Target Radiation Damage Lifetime

The MINOS near detector documented a gradual change in the NuMI neutrino spectrum over 3 years of running, which was attributed to target radiation damage; the neutrino yield was restored when a new target was installed. The neutrino rate decreased by around 10% in the peak, and increased slightly in the high energy tail. During this time, $6\times10^{20}$ protons on target were delivered at 120 GeV, for a total integrated beam power of 4.44 MW-month, and integrated proton flux of $8\times10^{21}$ POT/cm$^2$ at the center of the target. If one changed targets at LBNE after the same integrated beam power, and one assumes Project X will deliver 2.3 MW with 70% uptime, 4.4 targets / year would be required.

However, the spot size at LBNE will be larger, which should slow down the radiation damage. If one assumed that radiation damage scaled with proton flux density, i.e., (radius)$^{-2}$, then 2.4 targets / year would be needed for a 1.5 mm RMS beam spot. A somewhat more refined estimate would be that radiation damage scales by displacements per atom (DPA), which in this range scales more like (radius)$^{-3/2}$ at the maximum energy density point in the shower according to the MARS Monte Carlo, and 2.7 targets per year would be needed.

Radiation damage is predicted to be twice as fast for 60 GeV protons at the same beam power.

For comparison, in 4½ years of operation, there have been a total of 10 accesses into the NuMI target pile. A target swap for the NuMI beam-line takes about 2 weeks.

The above scaling motivates an LBNE design that provides a shorter target-swap down-time, or a larger beam spot to slow radiation damage, or a different target material that is more radiation hard. A "Gatling gun" target, such as the CNGS target, which can remotely rotate in fresh target material, would be an excellent fit for a Superbeam laid out like CNGS or NuMI/NoVA, where the target is upstream of the horns; unfortunately for LBNE the target needs to fit inside the first focusing horn.

# DECAY PIPE

## Optimization of Decay Volume Size

The LBNE nominal decay pipe length is 250 m, which provides approximately one half-life flight time for pion parents of 2.7 GeV neutrinos. The nominal diameter is 4 m. Figure 2 shows how neutrino yield varies with volume. The large amount of shielding required around the decay volume at FNAL (as at JPARC) makes increasing this volume quite expensive.

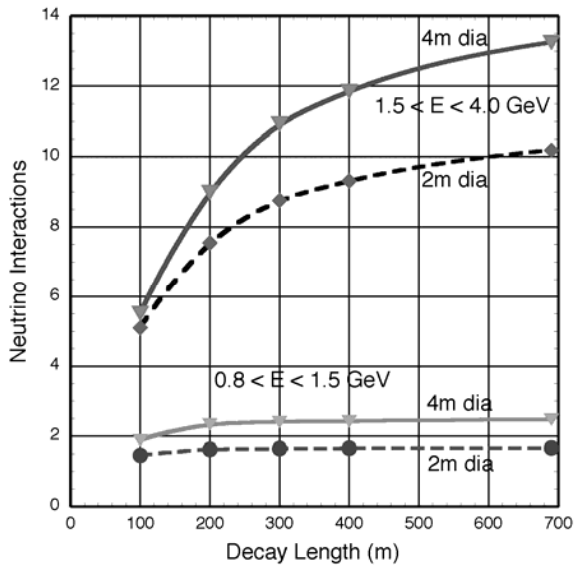

**FIGURE 2.** Relative expected neutrino interaction rate in LBNE detector as a function of decay volume length and radius. Solid (dashed) lines are for 4 m (2m) diameter decay volumes. Upper two curves are for the neutrino energy range around 1$^{st}$ oscillation maximum, lower two curves are for region relevant to the 2$^{nd}$ oscillation maximum.

## Selection of Decay Pipe Technology

Designing the decay volume hardware requires a tradeoff between risk, cost, and neutrino production efficiency. The energy deposited in the decay volume walls is projected to be 0.4 to 0.5 MW for 2 MW LBNE beam, which is high enough to require active cooling. After 30 days of running LBNE at 2 MW, the residual radiation on contact with a steel decay pipe wall would be (150 / 35 / 9) mSv/hr or (15,000 / 3,500 / 900) mrem/hr for ( 1 day / 1 month / 1 year ) of cool-down. Thus access by personnel for repair is impossible. The previous generation of neutrino beam-lines also have this non-repairability risk. NuMI, for example, has 5 miles of buried decay pipe water-cooling tubing that cannot be accessed. The T2K decay volume was built with 40 separate water cooling paths to reduce this risk.

*Vacuum with Water Cooling*

Building a vacuum vessel with water cooling yields the most neutrinos, as there is no gas to absorb or multiple scatter the pions. Vacuum was used for CNGS and for early NuMI running.

For vacuum in such a large volume, the stored energy is huge, and personnel protection is a serious issue. The window at the upstream end must be thin to minimize pion absorption, and is thus problematic. For NuMI, unexpected radiation-accelerated corrosion was discovered on this window, which prompted filling the decay pipe with helium to reduce the likelihood of failure and the stored energy that could be released in case of failure. T2K eliminated this window by making the target pile and decay volume a single helium volume; the entire volume is pulled to vacuum before the helium fill.

*Helium-Filled with Water Cooling*

A helium-filled decay volume yields several percent fewer neutrinos than vacuum.

The T2K scheme of also putting the target pile in the helium volume has the advantage of reducing corrosion of the horns and associated hardware. However, they must dump the helium inventory for access to repair/replace the target or horns.

*Air-Filled with Recirculating Air Cooling*

An air-filled decay volume will produce 10% less neutrino yield than helium-filled. However, it opens the possibility of re-circulating air through the volume itself to provide the cooling, rather than adding water-cooling channels. Preliminary calculations indicate the heat transfer from the pipe to air is good enough, with a flow of ~1,500 m$^3$ / minute required. NuMI currently uses such a recirculating air system to cool the target pile shielding; the LBNE decay volume system would be about twice the air flow rate and three times the heat load as NuMI, so it is eminently doable.

The main advantage is that all air equipment is external to the main radiation shielding, where it can be maintained after a few-hour beam-off cool-down period; there are no inaccessible water pipes.

Another advantage is that dehumidifiers in the air stream could collect a substantial fraction of beam-produced tritium before it can migrate somewhere else.

Two downsides of the air-cooling system are that substantial space is required underground for the air-handling equipment, and short-lived isotopes are produced in the air by the beam. Although it is a recirculating system, there will be some air leakage, particularly as outside air pressure changes, unless one builds an expensive and (as for helium) hard to repair totally sealed volume. A steady exhaust from the target hall area above the shielding is used in NuMI to keep air leaks from going in undesired directions. Based on experience from NuMI, for LBNE an external contained space of ~ 10,000 m$^3$ would be needed for decay of radio-activated air before release.

The existing NuMI facility could provide this space for LBNE.

## TRITIUM

Tritium is produced in hadronic showers, proportional to beam power, with cross sections not hugely sensitive to material choice, and hence initially nearly entirely embedded in the radiation shielding. NuMI produces a few hundred Ci/yr; a Superbeam will produce a few thousand Ci/yr. When NuMI turned on, we were surprised at how fast the tritium was getting out of the shielding, and added significant dehumidification capacity to intercept the majority of HTO before it could precipitate into the sump water. Over the last few years, of order 10% of the tritium produced in the shielding has been collected in the dehumidification condensate. The dehumidification systems have reduced the tritium in the ~600 liter/minute sump water stream by an order of magnitude and put the tritium in an ~ 0.2 liter/minute condensate waste stream. The tritium concentration in this condensate, which requires separate disposal, is around 100,000 pCi/ml when beam is running, and declines gradually after the beam is turned off, but concentrations remain elevated for months (and presumably years) later. For comparison, the limit for drinking water (U.S.) is 20 pCi/ml.

Tritium is not a show-stopper for a Superbeam, but needs to be carefully considered in the design. A NuMI-style dehumidification system could presumably be even more effective in a facility designed with Tritium mitigation in mind.

## CONCLUSION

Planning for mega-watt proton sources for neutrino beam-lines is underway, and superbeams could exist in about a decade. What each Superbeam looks like depends on the physics one wants to do. Once built, each will have limited flexibility (unless pre-designed and paid for). The target is the component where the current technology is pushed to the edge of the material property limit.

For JPARC and FNAL beams, by scaling from current targets, conventional solid targets appear very plausible, although detailed design and engineering remains to be done.

For T2K, the target hall / decay pipe / absorber for a Superbeam already exist. For others, significant design choices still remain.

## ACKNOWLEDGMENTS

The Monte Carlo calculations for LBNE beam of neutrino production rate, residual radiation, and heat deposition are the work of Byron Lundberg. Valeri Garkusha's group at IHEP, Protvino provided the design and calculations for a metal tube encapsulated graphite target. The Monte Carlo calculation of target DPA is due to Nikolai Mokhov. Kris Anderson supplied the engineering calculations for air cooling of the decay pipe. I wish to express thanks to the MINOS collaboration for the Near Detector neutrino spectrum information, and my continuing gratitude to many colleagues of CNGS, K2K, MiniBooNE, NuMI, T2K and WANF neutrino beam-lines from whom I have learned much.

This work is supported in part by Fermi Research Alliance, LLC under Contract No. DEAC02-07CH11359 with the United States Department of Energy.